\documentclass[a4paper,12pt]{article}

\usepackage{amssymb}
\usepackage[dvips]{graphicx}
\makeatletter
  
  \@addtoreset{equation}{section}
\makeatother
\makeatletter
\def\footnoteref#1{\def\@thefnmark{\ref{#1}}%
 \@footnotemark}
\makeatletter

\begin{document}

\begin{flushright}
OU-HET 370\\
hep-th/0011158\\
November 2000
\end{flushright}
\vspace*{1.5cm}
\begin{center}
{\Large {\bf Nonlinear Realization of Lorentz Symmetry}} \\
\bigskip
Naoto Yokoi\footnote{yokoi@het.phys.sci.osaka-u.ac.jp}\\
\bigskip
{\small
Department of Physics,\\
Graduate School of Science, Osaka University,\\
Toyonaka, Osaka 560-0043, JAPAN
}
\end{center}
\bigskip
\bigskip
\bigskip

\begin{abstract}
We explore a nonlinear realization of the (2+1)-dimensional 
Lorentz symmetry with a constant vacuum expectation value
of the second rank anti-symmetric tensor field. 
By means of the nonlinear realization, 
we obtain the low-energy effective action of the Nambu-Goldstone
bosons for the spontaneous Lorentz symmetry breaking. 
\end{abstract}

\newpage

\section{Introduction}
Field theories on the space-time with non-commutative coordinates
(non-commutative field theories) have been extensively studied 
for a few years \cite{Connes, IKKT, Seiberg-Witten}. 
A construction of the non-commutative field theories  
has been developed in \cite{Seiberg-Witten}: 
the world volume theory of Dp-brane with
a constant background NS-NS B-field is equivalent to a (p+1)-dimensional
non-commutative field theory whose non-commutative (constant) parameter
$\theta^{ij}$ is given by the background B-field $B_{ij}$.

It is a well-known fact that the theory with a constant second-rank
anti-symmetric tensor cannot have explicit Lorentz invariance
in (p+1)-dimensions for $\textrm{p} \geq 2$.
In string theory, NS-NS B-field is a dynamical 
field and its constant background field
can be regarded as the vacuum expectation value of the B-field.
In this view point, Lorentz symmetry is spontaneously broken
by the vacuum expectation value of the second rank anti-symmetric
tensor field.

One can ask naturally how Lorentz symmetry is realized in 
the broken theory and 
what is the Nambu-Goldstone boson for the spontaneous Lorentz
symmetry breaking. In this paper, we study the first problem.
We also discuss the low-energy effective action of the corresponding 
Nambu-Goldstone bosons, which is obtained from only the symmetry
argument.
The second problem, which is model-dependent, 
will be studied in the forthcoming paper \cite{H-Y}.

This paper is organized as follows. In section 2, we summarize
the nonlinear realization of general space-time symmetry.
In section 3, we study the nonlinear realization of the Lorentz symmetry
and construct the effective action of Nambu-Goldstone bosons 
and other fields invariant under the nonlinear transformation of 
the Lorentz symmetry.
In section 4, we discuss the related topics.  
 
\section{Nonlinear realization of space-time symmetry}
It is well-known that if a symmetry is broken spontaneously, the
broken symmetry is realized nonlinearly in the effective theory.
In this section we summarize the general theory of 
the nonlinear realization of space-time symmetry,
such as conformal symmetry, supersymmetry and so on, following
\cite{Ogievetsky} (references therein). 
We will apply this formalism to the Lorentz symmetry in the next section.

Let $G$ be a group of a space-time symmetry 
and $H$ be its stability subgroup, i.e., unbroken subgroup\footnote{
We denote the corresponding Lie algebra as ${\cal G}$ and ${\cal H}$ 
respectively.}. 
Then we assume that the generators satisfy the following commutation
relations:
\begin{eqnarray}
\left[ V_{i}, V_{j} \right] &=& f_{ijk} V_{k}
\qquad (i=1,\cdots,\textrm{dim}H), \\
\left[ V_{i}, Z_{a} \right] &=& f_{iab} Z_{b}
\qquad (a=1,\cdots,\textrm{dim}(G/H)), \label{reductivity}\\
\left[ Z_{a}, Z_{b} \right] &=& f_{abc} Z_{c} + f_{abi} V_{i},
\end{eqnarray}
where $V_{i} \in {\cal H},~Z_{a} \in {\cal G-H}$ and $f_{ABC}$ is the
structure constant of $G$. 
If $f_{abc} = 0$, the coset $G/H$ is called symmetric space.

From the above algebra 
we can construct a nonlinear realization of $G$ following the standard
procedure \cite{Weinberg}.
There is one peculiarity, however, caused by the fact 
that $G$ is a space-time symmetry.
Since $G$ is a space-time symmetry, $G$
necessarily includes translation generators.
According to the formalism in \cite{Ogievetsky}, 
in the case of nonlinear realization
of space-time symmetry, one should take the translation generators
as the ``broken'' generators $Z_{a} \in {\cal G-H}$,
even when these symmetries are not broken.
One reason is that the generators of $G$ must satisfy reductivity
condition Eq. (\ref{reductivity}) and the other is that the
coordinate dependence of the Nambu-Goldstone field (NG-field)
need to be properly transformed under $G$.

Considering this fact, we define the representative of the right coset 
$G/H$:
\begin{eqnarray}
L(x, \xi_{a}(x)) = e^{i x_{\mu} P^{\mu}} e^{i \xi_{a}(x) Z^{a}},
\end{eqnarray}
where $P^{\mu}~(\mu = 0,1,\cdots, D\textrm{--}1)$ are the unbroken 
translation generators and $\xi_{a}(x)$ is the NG-field
of the broken generator $Z^{a}$.
It is worth noting that the unbroken translation generators 
$P^{\mu}$ occupies a special place in $L(x,\xi_{a}(x))$.
Then we define the action of $G$ on the NG-fields as the left action:
\begin{eqnarray}
L(x^{'}, \xi_{a}^{'}(x^{'})) = g L(x, \xi_{a}(x)) = e^{i x^{'}_{\mu} P^{\mu}} 
e^{i \xi^{'}_{a}(x^{'}) Z^{a}} e^{i h_{i}(\xi_{a}(x), g) V^{i}},
\end{eqnarray}
where $g \in G$.\footnote{We will discuss only the global
symmetry where the transformation parameter is constant.}
One can see easily this action reproduces the transformation of
$\xi_{a}(x)$ under the translation, if we choose $g = e^{i a_{\mu} P^{\mu}}$:
\begin{eqnarray}
x^{'}_{\mu} = x_{\mu} + a_{\mu}, \quad \textrm{and} \quad 
\xi^{'}_{a}(x^{'}) = \xi_{a}(x).
\end{eqnarray}      
In general, under this $G$-action the NG-field $\xi_{a}(x)$ is transformed
nonlinearly, $\xi_{a}(x) \rightarrow \xi^{'}_{a}(x^{'})$.
One can define the $G$-action on other fields $\psi(x)$ 
which belong to a linear representation of $H$ as follows:
\begin{eqnarray}
\psi(x) \rightarrow \psi^{'}(x^{'}) = D \left(e^{i h_{i}(\xi_{a}(x),g)
V^{i}} \right) \psi(x),
\end{eqnarray}
where $g \in G$ and $D$ means the matrix representation of $H$  
which $\psi(x)$ belongs to. This is a nonlinear realization of the
space-time symmetry of $G$.

To construct the effective action invariant under the
$G$-action, we must obtain the quantities that transform covariantly
using the NG-fields $\xi_{a}(x)$.
Following the standard recipe \cite{Weinberg, Ogievetsky}, 
we define the Maurer-Cartan 1-form of
$G$:\footnote{$d = dx^{\mu} \partial/\partial x^{\mu}$ acts also on 
$\xi_{a}(x)$ implicitly.}
\begin{eqnarray}
L^{-1}dL(x,\xi_{a}(x)) = i {\cal D} x_{\mu} P^{\mu} + i {\cal D}
\xi_{a}(x) Z^{a} + i {\cal D} h_{i}(x) V^{i}, \label{cartan}
\end{eqnarray}
where 
\begin{eqnarray}
{\cal D}x^{\mu} &=& W(\xi_{a}(x))^{\mu}{}_{\nu} dx^{\nu}, \label{vielbein} \\
{\cal D} \xi_{a}(x) &=& {\cal D}_{\mu} \xi_{a}(x) dx^{\mu},
\qquad {\cal D} h_{i}(x) = {\cal D}_{\mu} h_{i}(x) dx^{\mu}.
\end{eqnarray}
Here, since the commutators between $P^{\mu}$ and $Z^{a}$ are nonzero
in general, nontrivial ``vielbein'' depending on the NG-fields $\xi_{a}(x)$ 
appears in (\ref{vielbein}).
Each term of the right hand side of Eq. (\ref{cartan}) 
transforms under the $G$-action as follows:
\begin{eqnarray}
{\cal D} x_{\mu}^{'} P^{\mu} &=& e^{i h_{i}(\xi_{a}(x), g) V^{i}} \left(
{\cal D} x_{\mu} P^{\mu} \right) e^{-i h_{i}(\xi_{a}(x), g) V^{i}} \\
{\cal D} \xi_{a}^{'}(x^{'}) Z^{a} &=& e^{i h_{i}(\xi_{a}(x), g) V^{i}} \left(
{\cal D} \xi_{a}(x) Z^{a} \right) e^{-i h_{i}(\xi_{a}(x), g) V^{i}} \\  
{\cal D} h_{i}^{'}(x^{'}) V^{i} &=& e^{i h_{i}(\xi_{a}(x), g) V^{i}} \left(
{\cal D} h_{i}(x) V^{i} \right) e^{-i h_{i}(\xi_{a}(x), g) V^{i}} \\
&& - i e^{i h_{i}(\xi_{a}(x), g) V^{i}} d e^{-i h_{i}(\xi_{a}(x), g) V^{i}}
\end{eqnarray}  
One can define the ``covariant derivative'' of the NG-fields 
using these quantities:
\begin{eqnarray}
\nabla_{\mu}\xi_{a}(x) = \frac{{\cal D}{\xi_{a}(x)}}{{\cal D} x^{\mu}}
= \left( W^{-1}(\xi_{a}(X)) \right)^{\nu}{}_{\mu} {\cal
D}_{\nu} \xi_{a}(x).
\end{eqnarray}
One can also define the ``covariant derivative'' of other fields
similarly:
\begin{eqnarray}
\nabla_{\mu}\psi(x) = \frac{{\cal D}{\psi(x)}}{{\cal D} x^{\mu}}
= \left( W^{-1}(\xi_{a}(x) \right)^{\nu}{}_{\mu} 
\left( \partial_{\nu} \psi(x) + i {\cal D}_{\nu} h_{i}(x) D\left( V^{i} 
\right) \psi(x) \right), \label{derivative of psi}
\end{eqnarray}
where $D\left( V^{i} \right)$ is a matrix representation of $H$ 
which $\psi(x)$ belongs to.

Finally we construct the $G$-invariant effective action 
by means of these covariant quantities.
In order to construct the $G$-inavriant effective action, one needs 
the $G$-invariant integration measure. If $H$ contains 
the $D$-dimensional homogeneous Lorentz group, 
one simply obtains
the $G$-invariant measure by the following replacement: 
\begin{eqnarray}
dx^{0} \wedge dx^{1} \wedge \cdots \wedge d x^{D-1} \longrightarrow
&& {\cal D} x^{0} \wedge {\cal D} x^{1} \wedge \cdots 
\wedge {\cal D} x^{D-1} \\ 
&& = \left( \det W \right) dx^{0} \wedge dx^{1} \wedge 
\cdots \wedge d x^{D-1}.
\end{eqnarray}
Then the resulting $G$-invariant effective action is given by
\begin{eqnarray}
S_{\textrm{eff}}(\xi_{a}(x),~\psi(x)) = \int d^{D}x \left( \det W \right) 
f(\nabla_{\mu}\xi_{a}(x),~\nabla_{\mu}\psi(x),~\psi(x)),
\end{eqnarray}
where $f$ is a $H$-invariant function.    

\section{Nonlinear realization of the Lorentz symmetry}
In this section, as a simplest example, 
we consider the nonlinear realization of the Lorentz symmetry in
(2+1)-dimensions. In (2+1)-dimensions, the Poincar\'e algebra ($\sim
iso(2,1)$) is given by the Lorentz generators $M_{\mu \nu}$ and 
the translation generators $P_{\mu}$:\footnote{$\eta_{\mu \nu}=\eta^{\mu
\nu} = \textrm{diag.}(+1,-1,-1)$ and $\epsilon^{012}= +1$.}
\begin{eqnarray}
\left[ P^{\mu}, P^{\nu} \right] &=& 0, \hspace{2.6cm} \left[ J^{\mu}, 
P^{\nu} \right] = -i \epsilon^{\mu \nu \rho} P_{\rho}, \\  
\left[ J^{\mu}, J^{\nu} \right] &=& -i \epsilon^{\mu \nu \rho} J_{\rho}, 
\qquad \textrm{where}~~J^{\mu} \equiv \frac{1}{2} \epsilon^{\mu \nu
\rho} M_{\nu \rho}. \qquad (\mu = 0,1,2)
\end{eqnarray}
The homogeneous Lorentz subgroup 
generated by $J^{\mu}$ forms $SO(2,1) \sim SL(2, {\bf R})$
and we take a basis as 
\begin{eqnarray}
J^{0} = \frac{1}{2} \sigma_{3}, \quad J^{1} = \frac{i}{2} \sigma_{1},
\quad \textrm{and}~~ J^{2} = \frac{i}{2} \sigma_{2},
\end{eqnarray} 
where $\sigma_{i}$'s are Pauli matrices.

We consider the situation that a 
second rank anti-symmetric tensor field $B_{\mu \nu}$
has a nonzero vacuum expectation value (vev).
We assume only the component $B_{12}$ has a constant 
nonzero vev\footnote{If the vev of $\tilde{B}^{\mu} = \frac{1}{2}
\epsilon^{\mu \nu \rho} B_{\nu \rho}$ is a time-like constant vector, 
one can transform it to this form by the Lorentz transformation.}, i.e.,
\begin{eqnarray}
\langle B_{12} \rangle = B \neq 0~~(= \textrm{const.}), \quad
\textrm{and} \quad \langle B_{0i} \rangle = 0. \label{vev}
\end{eqnarray}
One can easily find the boost generators $M_{01}$ and $M_{02}$ or
$J^{1}$ and $J^{2}$ are broken by the vev of $B_{\mu \nu}$, (\ref{vev}). 
However, since the vev is a constant, the translation generators
$P^{\mu}$ are not broken. Thus, in this case, $G$ is the Poincar\'e group
($ISO(2,1)$) and $H$ is the subgroup generated by the translations and
the spatial rotation. Therefore one may expect naively 
the nonlinear realization of the Lorentz symmetry can be 
simply constructed from the coset $G/H \sim
SO(2,1)/SO(2) \sim SL(2,{\bf R})/U(1)$.

However, according to the discussion in the previous section, 
we should decompose ``broken'' and ``unbroken'' generators in 
a different manner:
\begin{eqnarray}
G~:~ISO(2,1),&& H~:~U(1) \nonumber \\
\textrm{``Unbroken generator''} &:&J^{0} \quad 
\hspace*{3.2cm} \in {\cal H}, \nonumber \\
\textrm{``Broken generator''} &:&J^{1},~J^{2},~P^{\mu} \hspace{2cm}
\in {\cal G-H}.
\end{eqnarray}
Using this decomposition, the representative of $G/H$ is defined as
follows:
\begin{eqnarray}
L(x, \phi_{a}(x)) &=& e^{i x_{\mu} P^{\mu}} e^{i \phi_{a}(x) J^{a}}
= e^{i x^{0} P_{0} + i \bar{z} P_{+} + i z P_{-}} e^{i
\bar{\phi}(x) J^{+} + i \phi(x) J^{-}}, \label{representative} \\
\textrm{where} && z = x^{1} + i x^{2}, \quad P_{\pm} = 
\frac{1}{2} \left(P_{1}\pm i P_{2}\right), \\
&& \hspace{-0.6cm}\phi(x) = \phi_{1}(x)+ i \phi_{2}(x),
\quad J^{\pm} = \frac{1}{2} \left(J^{1}\pm i J^{2}\right).
\end{eqnarray}
$\phi_{1}(x)$ and $\phi_{2}(x)$ are the NG-bosons for the broken
generators $J^{1}$ and $J^{2}$ respectively. The candidates of
$\phi_{a}(x)$ in specific models will be discussed in the next section.
The $G$-action is defined as the left-action:
\begin{eqnarray}
L^{'}(x^{'}, \phi_{a}^{'}(x^{'})) = g L(x, \phi_{a}(x)). \qquad (g
\in G) \label{g-action}
\end{eqnarray}
In particular, one can find that this $G$-action properly 
reproduces the linear transformation under the unbroken symmetries.
If $g = e^{i a_{\mu} P^{\mu}}$, the transformation becomes
\begin{eqnarray} 
x_{\mu}^{'} = x_{\mu} + a_{\mu}, \quad \phi^{'}(x^{'}) = \phi(x),
\quad \bar{\phi}^{'}(x^{'}) = \bar{\phi}(x).
\end{eqnarray}
And under the unbroken spatial rotation, i.e., $g = e^{i b J^{0}}~(= e^{i b
M_{12}})$, the transformation becomes
\begin{eqnarray}
x^{0 '} = x^{0}, \quad z^{'} = e^{-ib} z, \quad \phi^{'}(x^{'}) =
e^{-ib} \phi(x), \quad \bar{\phi}^{'}(x^{'}) = e^{ib} \bar{\phi(x)}.
\end{eqnarray}

We can obtain the Maurer-Cartan 1-form explicitly using the
representative (\ref{representative}). 
\begin{eqnarray}
L^{-1}dL(x,\phi_{a}(x)) &=& e^{-i \bar{\phi}(x)J^{+}-i \phi(x) J^{-}}
\left( i P_{0} dx^{0} + i P_{+} d\bar{z} + i P_{-} dz \right) 
e^{i \bar{\phi}(x)J^{+}+i \phi(x) J^{-}} \nonumber \\
&& +~e^{-i \bar{\phi}(x)J^{+}-i \phi(x) J^{-}} 
d \left( e^{i \bar{\phi}(x)J^{+}+i \phi(x) J^{-}} \right) \\
&\equiv& i {\cal D}x^{\alpha} P_{\alpha} + i {\cal D} \bar{\phi}(x)
J^{+} + i {\cal D} \phi(x) J^{-} + i {\cal D} h(x) J^{0},
\end{eqnarray}
where $x^{\alpha} = (x^{0}, z, \bar{z})$.
The explicit form is given by
\begin{eqnarray}
{\cal D}x^{\alpha} = W^{\alpha}{}_{\beta}(\phi, \bar{\phi}) 
dx^{\beta}, \quad 
{\cal D} \phi(x) = {\cal D}_{\alpha} \phi(x) dx^{\alpha}, \quad
{\cal D} h(x) = {\cal D}_{\alpha} h(x) dx^{\alpha},  
\end{eqnarray}
where
\begin{eqnarray}
W^{\alpha}{}_{\beta}(\phi, \bar{\phi})= 
\left( 
\begin{array}{ccc}
\cosh(|\phi|) & -\frac{1}{2}i\bar{\phi}\frac{\sinh(|\phi|)}{|\phi|} & 
\frac{1}{2}i\phi\frac{\sinh(|\phi|)}{|\phi|} \\
i\phi\frac{\sinh(|\phi|)}{|\phi|} & 
\frac{1 + \cosh(|\phi|)}{2} &
\frac{1}{2}\phi^2 \left(\frac{1 - \cosh(|\phi|)}{|\phi|^2}\right) \\
-i\bar{\phi}\frac{\sinh(|\phi|)}{|\phi|} & 
\frac{1}{2}\bar{\phi}^2 \left(\frac{1 - \cosh(|\phi|)}{|\phi|^2}\right) &
\frac{1 + \cosh(|\phi|)}{2} \\
\end{array}
\right)
\end{eqnarray}
and 
\begin{eqnarray}
{\cal D}_{\alpha} \phi(x) &=& \frac{1}{2}\left( \frac{\phi
\partial_{\alpha} \bar{\phi} + \bar{\phi}
\partial_{\alpha} \phi}{\bar{\phi}}-\frac{(\phi
\partial_{\alpha} \bar{\phi} - \bar{\phi} \partial_{\alpha} \phi) 
\sinh(|\phi|)}{\bar{\phi}|\phi|} \right) \label{cartan of phi}\\ 
{\cal D}_{\alpha} h(x) &=&
\frac{\sinh^{2}\left(\frac{|\phi|}{2}\right)}{i~|\phi|^{2}} \left(\phi
\partial_{\alpha} \bar{\phi} - \bar{\phi} \partial_{\alpha} \phi \right),  
\end{eqnarray}
and ${\cal D}_{\alpha} \bar{\phi}(x) = \overline{{\cal D}_{\alpha}
\phi(x)}$, denoting the complex conjugation of (\ref{cartan of phi}).

Following the recipe explained in the previous section, one can
obtain the covariant derivative of the NG-fields $\phi(x)$ and
$\bar{\phi}(x)$.
\begin{eqnarray}
\nabla_{\alpha} \phi(x) = \left( W^{-1} \right)^{\beta}{}_{\alpha}
{\cal D}_{\beta} \phi(x), \quad \nabla_{\alpha}\bar{\phi}(x) 
= \left( W^{-1} \right)^{\beta}{}_{\alpha} {\cal D}_{\beta}
\bar{\phi}(x),
\end{eqnarray}
where
\begin{eqnarray}
\nabla_{0}\phi &=& e^{i \theta} \cosh(\rho)(\partial_{0} \rho + i \sinh(\rho)
\partial_{0} \theta) -i e^{2 i \theta} \sinh(\rho) (\partial_{z} \rho 
+ i \sinh(\rho) \partial_{z}\theta) \nonumber \\ &&
+ i \sinh(\rho) (\partial_{\bar{z}} \rho + 
i \sinh(\rho) \partial_{\bar{z}}\theta) \\
\nabla_{z}\phi &=& \frac{i}{2} \sinh(\rho)
(\partial_{0} \rho + i \sinh(\rho)
\partial_{0} \theta) + e^{i \theta} \cosh^{2}(\frac{\rho}{2}) 
(\partial_{z} \rho + i \sinh(\rho) \partial_{z}\theta) \nonumber \\ &&
- e^{-i \theta} \sinh^{2}(\frac{\rho}{2}) 
(\partial_{\bar{z}} \rho + i \sinh(\rho) \partial_{\bar{z}
}\theta) \\
\nabla_{\bar{z}}\phi &=&-\frac{i}{2} e^{2 i \theta} \sinh(\rho)
(\partial_{0} \rho + i \sinh(\rho)
\partial_{0} \theta) - e^{3 i \theta} \sinh^{2}(\frac{\rho}{2})
 (\partial_{z} \rho + i \sinh(\rho) \partial_{z}\theta) \nonumber \\ &&
+ e^{i \theta} \cosh^{2}(\frac{\rho}{2}) 
(\partial_{\bar{z}} \rho + i \sinh(\rho) \partial_{\bar{z}
}\theta),
\end{eqnarray}
and $\nabla_{0}\bar{\phi}(x) = \overline{\nabla_{0} \phi (x)}$,
$\nabla_{z} \bar{\phi}(x) = \overline{\nabla_{\bar{z}}\phi(x)}$, 
$\nabla_{\bar{z}} \bar{\phi}(x) = \overline{\nabla_{z}\phi(x)}$.
Here we have introduced $\rho(x)$ and $\theta(x)$ by 
$\phi(x) \equiv \rho(x) e^{i \theta(x)}$ for simplicity. 
In order to construct the covariant derivative of 
other fields $\psi(x)$ from Eq. (\ref{derivative of psi}), 
we need the explicit form of 
\begin{eqnarray}
\nabla_{\alpha}h(x) = \left( W^{-1} \right)^{\beta}{}_{\alpha} 
{\cal D}_{\beta} h(x), \label{derivative of h}
\end{eqnarray}
namely
\begin{eqnarray}
\nabla_{0}h(x) &=& -2 \sinh^{2} (\frac{\rho}{2} ) \left( \cosh(\rho) 
\partial_{0} \theta - i e^{i \theta} \sinh(\rho) \partial_{z} \theta
+ i e^{-i \theta} \sinh(\rho) \partial_{\bar{z}} \theta \right) \\
\nabla_{z} h(x) &=& -2 \sinh^{2} (\frac{\rho}{2} ) 
\left(\frac{i}{2} e^{-i \theta} \sinh (\rho) \partial_{0} \theta 
+ \cosh^{2} (\frac{\rho}{2} ) 
\partial_{z} \theta \right. \nonumber \\ && \left. - e^{-2i \theta} 
\sinh^{2} ( \frac{\rho}{2} ) \partial_{\bar{z}} \theta \right)
\end{eqnarray}
and $\nabla_{\bar{z}}h(x) = \overline{\nabla_{z}h(x)}$. 
For example, if $\psi(x)$ is a real scalar field, the covariant derivative 
is given by (\ref{derivative of psi}) with $D(V^{i})=D(J^{0})=0$, i.e.,
\begin{eqnarray}
\nabla_{\alpha} \psi(x) = \left( W^{-1} \right)^{\beta}{}_{\alpha}
\partial_{\beta} \psi(x),
\end{eqnarray}
whose explicit forms are
\begin{eqnarray}
\nabla_{0}\psi(x) &=& \left( \cosh(\rho) \partial_{0} -i e^{i \theta}
\sinh(\rho) \partial_{z} + ie^{-i \theta} \sinh(\rho)
\partial_{\bar{z}} \right) \psi(x), \\
\nabla_{z}\psi(x) &=& \left( \frac{i}{2} e^{-i \theta} \sinh(\rho) 
\partial_{0} + \cosh^{2} (\frac{\rho}{2} ) \partial_{z} 
- e^{-2 i \theta} \sinh^{2} (\frac{\rho}{2} ) \partial_{\bar{z}}
\right) \psi(x),
\end{eqnarray}
and $\nabla_{\bar{z}} \psi(x) = \overline{\nabla_{z}\psi(x)}$. 
Similarly, for a spinor field, the covariant derivative is
given by (\ref{derivative of psi}) with $D(J^{0})= \sigma_{3}/2$~:
\begin{eqnarray}
\nabla_{\alpha} \psi_{a}(x) = \left( W^{-1} \right)^{\beta}{}_{\alpha}
\partial_{\beta} \psi_{a}(x) + i \nabla_{\alpha} h(x) 
\left(\frac{\sigma_{3}}{2}\right)_{a}{}^{b} \psi_{b}(x).
\end{eqnarray}

As discussed in the previous section, the $G$-invariant integration
measure is also needed. In our case, $H$ does not
contain the (2+1)-dimensional homogeneous Lorentz group.
Hence, we have three $G$-invariant integration measures, which are
1-dimensional measure ${\cal D} x^{0}$, 2-dimensional measure
${\cal D} z \wedge {\cal D} \bar{z}$, and (2+1)-dimensional measure
${\cal D} x^{0} \wedge {\cal D} z \wedge {\cal D} \bar{z}$.
Because we want to obtain the (2+1)-dimensional $G$-invariant
effective action, we take the $G$-invariant measure 
${\cal D} x^{0} \wedge {\cal D} z \wedge {\cal D} \bar{z} =
(\det W)dx^{0} \wedge dz \wedge d\bar{z}$. 

Finally, the $G$-invariant effective action of NG-fields $\phi(x)$ and
$\bar{\phi}(x)$ is given by
\begin{eqnarray}
S_{\textrm{eff}}(\phi(x), \bar{\phi}(x)) = \int d^{3}x (\det W)~
f \left( \nabla_{\alpha} \phi(x), \nabla_{\alpha} \bar{\phi}(x) \right),
\end{eqnarray}
where $f$ is a $H$-invariant function. Thus we can write down explicitly 
the effective action up to two derivatives of NG-fields:
\footnote{
Although one may wonder if the second and third term in (\ref{effective
action}) are independent of each other, the difference of these terms is
a topological term, i.e., a nontrivial surface term and 
cannot be ignored in general.   
} 
\begin{eqnarray}
S_{\textrm{eff}}^{(2)}(\phi(x), \bar{\phi}(x)) 
&=& \int d^{3}x \left\{ a \nabla_{0}\phi \nabla_{0}\bar{\phi} + b \nabla_{z}
\bar{\phi} \nabla_{\bar{z}} \phi + c \nabla_{z} \phi \nabla_{\bar{z}}
\bar{\phi} \right.\nonumber \\ 
&& \hspace*{1.2cm} + d \left( (\nabla_{z} \phi)^{2} 
+ (\nabla_{\bar{z}}
\bar{\phi})^{2} \right) + i e \left((\nabla_{z} \phi)^{2} - 
(\nabla_{\bar{z}} \bar{\phi})^{2} \right) \nonumber \\
&& \hspace*{1.2cm} \left. + k (\nabla_{z} \phi +
\nabla_{\bar{z}} \bar{\phi} ) + i l (\nabla_{z} \phi-\nabla_{\bar{z}}
\bar{\phi}) \right\}. \label{effective action}
\end{eqnarray}
Here we make a comment: the Jacobian factor $\det W = 1$ identically
for this Lorentz symmetry, although this factor depends 
nontrivially on the NG-fields and 
contributes to the effective action for the 
conformal symmetry \cite{Salam-Strathdee2}, supersymmetry
\cite{Volkov-Akulov} and superconformal symmetry \cite{Uematsu}.  
This is because the
simple integration measure $d^3 x$ is already Lorentz invariant.
This $G$-invariant effective action of the NG-fields up to two
derivative terms has not only two derivative terms but 
also one derivative terms and seven undetermined parameters. 
This is quite different from 
the counterpart of internal symmetry, which includes only two
derivative terms and one parameter \cite{Weinberg}. 

In the same way, one can construct the $G$-invariant effective action
of other field $\psi(x)$: 
\begin{eqnarray}
S_{\textrm{eff}}(\psi(x)) = \int d^3 x~p \left(\psi(x),
\nabla_{\alpha} \psi(x) \right), \label{effective action of psi}
\end{eqnarray}    
where $\nabla_{\alpha} \psi(x)$ is given by (\ref{derivative of psi}) 
and (\ref{derivative of h}) and $p$ is a $H$-invariant function.
It is worth commenting that this action, even up to two derivatives of
fields, necessarily includes derivative couplings 
between the NG-fields $\phi(x)$ and \textit{any} other fields $\psi(x)$. 

Thus we have obtained the effective action invariant under the
nonlinear transformation (\ref{g-action}) of the Lorentz symmetry.
This gives the low-energy effective action of the NG-bosons for the 
spontaneous Lorentz symmetry breaking in (2+1)-dimensions.

\section{Discussions}
We have discussed the nonlinear realization of the Lorentz symmetry in 
(2+1)-dimensions and obtained the low-energy effective action 
invariant under the nonlinear transformation of the Lorentz symmetry.    

We comment shortly on the several examples that realize the 
spontaneous Lorentz symmetry breaking. The first example is gauge 
invariant field theory of the second rank anti-symmetric tensor 
field $B_{\mu \nu}$ \cite{Townsend, Kimura}.
In the free field theory, the vev of $B_{\mu \nu}$ can be
a non-zero constant, thus the vev (\ref{vev}) is realized. 
In the gauge fixed quantum theory\footnote{In order to discuss the spontaneous
Lorentz symmetry breaking, the gauge fixing condition should be
manifestly Lorentz invariant. Thus we take a covariant gauge, say,
the Feynman gauge.},
the NG-bosons for this spontaneous Lorentz symmetry breaking 
are $B_{01}$ and $B_{02}$. 
$\phi_{1}$ and $\phi_{2}$ in the previous section correspond to $B_{01}$ 
and $B_{02}$ respectively.
In perturbation theory, 
the low-energy effective action of 
$B_{01}$ and $B_{02}$
is given by
(\ref{effective action}) obtained in 
the previous section.
Unfortunately, in (2+1)-dimensions, $B_{\mu \nu}$ does not have the 
physical degree of freedom. 

The second example is the quantum electrodynamics with 
the Chern-Simons coupling in (2+1)-dimensions. 
In this theory, spontaneous magnetization occurs and thus 
spontaneous Lorentz symmetry breaking is realized 
\cite{Hosotani1, Hosotani2}.
In this case, the field strength of photon condenses and 
the vev's of $F_{\mu \nu}$ become 
\begin{eqnarray}
\langle F_{12} \rangle = B \neq 0~(= \textrm{const.}), \quad
\textrm{and} \quad 
\langle F_{0i} \rangle = 0.
\end{eqnarray}
Then the NG-bosons are $F_{01}$ and $F_{02}$\footnote{In the physical
Hilbert space, these two NG-bosons are not independent each other 
\cite{Hosotani2}.}, which correspond to $\phi_{1}$ and $\phi_{2}$ 
respectively.
Thus the low-energy effective action of 
$F_{01}$ and $F_{02}$
will be given by (\ref{effective action}) 
and the coupling of fermions and $F_{01}$ and $F_{02}$
is given by (\ref{effective action of psi}).       

The third example is the field theory of a second rank 
anti-symmetric tensor field coupled with an abelian gauge field.
This example will be extensively studied in the forthcoming paper \cite{H-Y}. 

The generalization to higher dimensions is 
straightforward. The construction discussed in this paper 
can be applied to the spontaneous Lorentz symmetry breaking 
in higher dimensions.
For more realistic example, the nonlinear realization of 
the Lorentz symmetry in (3+1)-dimensions on the vacuum
\begin{eqnarray}
\langle B_{12} \rangle = B \neq 0~(= \textrm{const.}),  \quad
\textrm{and~~~others}~= 0
\end{eqnarray}
can be similarly constructed and the corresponding low-energy 
effective action can also be constructed. 

The generalization to supersymmetric models is also 
interesting.

{\bf Acknowledgements} \\
The author is very grateful to K. Higashijima for 
the useful suggestions and discussions and 
careful reading of this manuscript.
The author also thanks Y. Hosotani for the useful discussions.

\end{document}